\documentstyle[twoside,fleqn,espcrc2,epsf]{article}

\title{B meson decay constant with the Wilson and Clover
  heavy quark actions\thanks{presented by S. Hashimoto}}

\author{JLQCD Collaboration : 
        S. Aoki\address{Institute of Physics, University of Tsukuba,
        Tsukuba, Ibaraki 305, Japan},
        M. Fukugita\address{Institute for Cosmic Ray Research,
        University of Tokyo, Tanashi, Tokyo 188, Japan},
        S. Hashimoto\address{Computing Research Center,
        High Energy Accelerator Research Organization (KEK),\\
        Tsukuba, Ibaraki 305, Japan},
        N. Ishizuka$^{\rm a,}$\address{Center for Computational Physics,
        University of Tsukuba, Tsukuba, Ibaraki 305, Japan},
        Y. Iwasaki$^{\rm a,d}$,\\
        K. Kanaya$^{\rm a,d}$
        Y. Kuramashi\address{Institute of Particle and Nuclear Studies,
        High Energy Accelerator Research Organization (KEK),
        Tsukuba, Ibaraki 305, Japan},
        M. Okawa$^{\rm e}$,
        A. Ukawa$^{\rm a}$,
        T. Yoshi\'e$^{\rm a,d}$
}

\begin{document}

\begin{abstract}

We present results of our quenched study of the $B$ meson decay 
constant obtained with a parallel set of simulations with the Wilson 
and Clover actions at $\beta$=5.9, 6.1 and 6.3. Systematic errors
associated with the large $b$-quark mass are analyzed within the 
Fermilab non-relativistic formalism.  As our best estimate in the continuum 
limit we obtain $f_B$=163$\pm$16 MeV and  $f_{B_s}$=175$\pm$18 MeV with the 
Clover action.

\end{abstract} 

\maketitle

\section{Introduction}

A reliable determination of the $B$ meson decay constant is a 
subject yet to be completed in lattice QCD.
We have been pursuing this goal employing the relativistic formalism for heavy 
quark.  Our results for the Wilson action has been reported in 
Refs.~\cite{JLQCD_heavy-light_95,JLQCD_heavy-light_96}.
Since Lattice'96 we have carried out a parallel set of simulations with the 
$O(a)$-improved Clover action.  We have analyzed the results for the 
two actions within the Fermilab non-relativistic 
formalism\cite{El-Khadra_Kronfeld_Mackenzie_97} with the view to 
understand the systematic error due to a large value of $b$-quark mass.
In this article we report a summary of results from the simulations and 
analyses.

\section{Simulation}

\begin{table}[htbp]
  \setlength{\tabcolsep}{0.3pc}
  \begin{center}
    \caption{Simulation parameters. The lattice scale quoted 
      is fixed by the string tension 
      $\protect\sqrt{\sigma}$=427~MeV\protect\cite{Bali_Schilling_92}.}
    \begin{tabular}{lllll}
      \hline
      action & $\beta$ & 5.9 & 6.1 & 6.3 \\
      \hline
             & size & 16$^3\times$40
                    & 24$^3\times$64
                    & 32$^3\times$80 \\
             & $1/a$ (GeV)& 1.60(1) & 2.29(1) & 3.05(2) \\
             & $L$ (fm)   & 2.0 & 2.1 & 2.1 \\
      \hline
      Wilson &$N_{conf}$  & 150   & 100   & 100 \\
      Clover &$N_{conf}$  & 540   & 200   & 166 \\
             &$c_{sw}$    & 1.580 & 1.525 & 1.484 \\
      \hline
    \end{tabular}
  \label{tab:parameters}
  \end{center}
\vspace*{-12mm}
\end{table}

The parameters of our simulations are listed in
Table \ref{tab:parameters}.  
For the clover coefficient we use
the tadpole-modified one-loop
value\cite{Luscher_Weisz_96} given by
$c_{sw}\!=\!P^{-3/4}[1\!+\!0.199\alpha_V(1/a)]$ with $P$ the average plaquette.
Heavy quarks are simulated for 7 values of the hopping parameter $\kappa$
covering the $c$ and $b$ quark masses, and 4 values of $\kappa$ 
are employed for light quark.

The heavy-light decay constant $f_P$ is extracted 
from the correlators of the axial vector current $A_4$ and
a smeared pseudoscalar density
$P^S(x)=\sum_{\vec{r}}\phi(|\vec{r}|)\bar{Q}(x+r)\gamma_5q(x)$
on the Coulomb gauge fixed gluon configurations, where 
$\phi(|\vec{r}|)$ is the pseudoscalar wave function measured 
for each heavy and light quark masses.

\begin{figure}[t]
  \begin{center}
    \vspace*{-5mm}
    \epsfxsize=65mm \epsffile{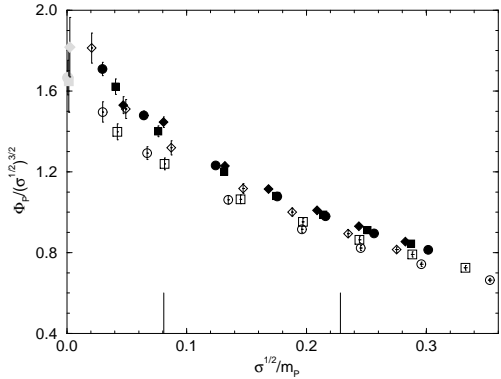}
    \vspace*{-9mm}
    \caption{$\Phi_P/\protect\sqrt{\sigma}^{3/2}$ as a function of 
      $\protect\sqrt{\sigma}/m_P$.
      Filled (open) symbols are Clover (Wilson) results.
      Circles, squares and diamonds correspond to values
      at $\beta$=5.9, 6.1 and 6.3.}
    \label{fig:heavy-light_decay_constant}
  \end{center}
\vspace*{-12mm}
\end{figure}

A new perturbative ingredient in our work is the 
recent one-loop result\cite{Aoki_etal_97} for 
the pole ($m_1^Q$) and kinetic ($m_2^Q$) masses of 
heavy quark and the renormalization factor $Z_A(am_Q)$ of the 
axial vector current for finite bare heavy quark mass $m_Q$ 
(see Refs.~\cite{Kronfeld_Mertens_93,Kuramashi_97} for previous
calculation for the Wilson case).
Effects of finite values of $am_Q$ in $Z_A$ is significant, 
reducing $f_B$ by 5--2\% for the Wilson action and increasing 
it by a similar magnitude for the Clover case 
compared to the value obtained with $Z_A(am_Q=0)$. 

We define the heavy-light meson mass 
by\cite{Bernard_Labrenz_Soni_94,JLQCD_heavy-light_96} 
$m_P\!=\!m_{P1}\!+\!m_2^{Q}\!-\!m_1^{Q}$
with $m_{P1}$ the measured meson pole mass and 
the one-loop perturbative result\cite{Aoki_etal_97} applied for 
$m_2^{Q}\!-\!m_1^{Q}$.
This definition does not have the problem of the measured kinetic mass
that the $b$ quark mass can not be determined
consistently from heavy-light and heavy-heavy mesons
\cite{SCRI_95,Kronfeld_96,JLQCD_heavy-light_96}.

\section{Results}

\begin{figure}[t]
  \begin{center}
    \vspace*{-5mm}
    \epsfxsize=65mm \epsffile{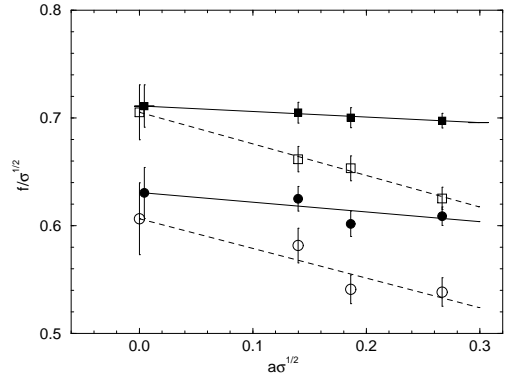}
    \vspace*{-9mm}
    \caption{Continuum extrapolation of $f_B$ (circles) and
      $f_D$ (squares).
      Filled (open) symbols are Clover (Wilson) results.}
    \label{fig:f_B_D_continuum}
  \end{center}
\vspace*{-12mm}
\end{figure}

We plot the quantity defined by 
$\Phi(m_P)\!=\!(\alpha_s(m_P)/\alpha_s(m_B))^{2/\beta_0}f_P\sqrt{m_P}$ 
in Fig.~\ref{fig:heavy-light_decay_constant} as a 
function of $1/m_P$ for the 
the Clover(filled symbols) and Wilson (open symbols) actions.
The light quark mass is linearly extrapolated to the chiral limit, and 
$\alpha_s(\mu)$ is calculated with the standard 2-loop definition 
with $\Lambda_{QCD}=$ 295 MeV.  We normalize the results by 
the string tension $\sigma$\cite{Bali_Schilling_92}
since we primarily wish to examine the question of large-$am_Q$ errors in 
this figure. 
Vertical lines indicate the position of 
$B$ and $D$ mesons for $\sqrt{\sigma}=427$~MeV. Data points at $1/m_P=0$ 
are the static results\cite{Duncan_et_al_95}, to which our results 
for the same set of $\beta$ values converge.

In Fig.~\ref{fig:f_B_D_continuum} we plot the continuum 
extrapolation of $f_B$ and $f_D$.  The Wilson 
results exhibit a scaling violation of 11-5\% in our range 
of lattice spacing $a^{-1}\approx$ 1.6--3~GeV, while the Clover 
results show a significantly reduced variation of 4--2\%.  These 
magnitudes are common to $f_B$ and $f_D$.  
Furthermore the continuum values obtained 
with the two actions by a linear extrapolation agree 
within the statistical error of about 5\%.

We emphasize that this agreement does not necessarily mean 
that systematic errors due to heavy quark are negligibly small.
In the non-relativistic interpretation, the equivalent Hamiltonian 
for Wilson-type actions has the form
\begin{equation}
  \label{eq:Non-relativistic_Hamiltonian}
  H = \bar{Q} \left[ m_1
    - \frac{\vec{D}^2}{2m_2}
    - \frac{i\vec{\sigma}\cdot\vec{B}}{2m_B}
    + O(1/m_Q^2) \right] Q.
\label{eq:nrqcd}
\end{equation}
For the Wilson action for which $m_B\ne m_2$, the leading error in 
the decay constant due to heavy quark is $O((c_B-1)\Lambda_{QCD}/m_Q)$
with $c_B\!=\!m_2/m_B$.
For the $B$ meson an examination $c_B$ at the tree level shows that 
a linear extrapolation of $c_B$ from our range of $a^{-1}$ 
leads to a value $|c_B-1|\!\approx$0.4 
at $a$=0.  We should therefore allow  
an $O(3\%)$ error unremoved in the continuum limit where we used 
$\Lambda_{QCD}$=0.3~GeV.
The same magnitude of error also remains for $f_D$.

There are two more sources of systematic error we need to consider. 
One is $m_Q$-independent scaling violation 
of $O(a\Lambda_{QCD})$, which we estimate to be 10\% at our smallest 
$a^{-1}\!\approx$3~GeV. The other is $O(\alpha_s^2)$ 
uncertainty due to the use of one-loop value for $Z_A$, which is $O(4\%)$ with 
$\alpha_s\approx 0.2$.  Adding all the errors by quadrature leads to  
a combined systematic error of $O(11\%)$ in the decay constant 
for the Wilson case.

For the Clover action for which $m_B=m_2$ to $O(\alpha_s)$, 
the large-$am_Q$ errors have the form
$O(\alpha_s\Lambda_{QCD}/m_Q)$ and $O(\Lambda_{QCD}^2/m_Q^2)$.  We estimate 
their magnitude to be $O(1\%)$ incorporating the effect of coefficients,  
similar to $c_B$, that vanish in the continuum limit. 
The scaling violation errors are $O(\alpha_sa\Lambda_{QCD},a^2\Lambda_{QCD}^2)$
which are small at $O(2\%)$.  
With the 2-loop error of $O(4\%)$ from $Z_A$ and 
an additional error of $O(2\%)$ arising from 
the field rotation ignored in the present calculation, 
the combined systematic error amounts to $O(5\%)$ for the Clover 
case.

\begin{figure}[t]
  \begin{center}
    \vspace*{-5mm}
    \epsfxsize=65mm \epsffile{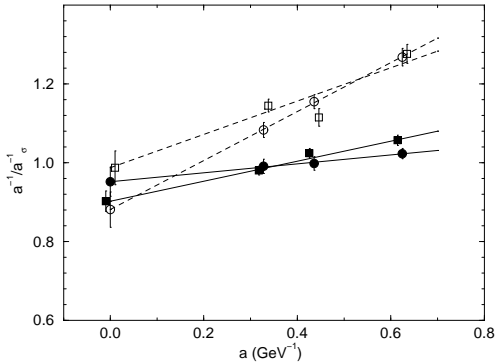}
    \vspace*{-9mm}
    \caption{Ratio of lattice scale obtained from $m_{\rho}$
      (circles) and from $f_{\pi}$ (squares) to that from the
      string tension.
      Filled (open) symbols are for Clover (Wilson) action.}
    \label{fig:lattice_scale_ratio}
  \end{center}
\vspace*{-15mm}
\end{figure}

So far we have used the string tension $\sigma$ to set the scale. 
In Fig.~\ref{fig:lattice_scale_ratio} we plot the ratio of 
$a^{-1}$ obtained with $m_\rho$ and $f_\pi$ to 
that with $\sigma$.
We use the variation of the ratio to estimate the uncertainty in setting
the scale, which we take to be 10\% for the Wilson case and 5\% for the 
Clover case.  This uncertainty includes possible quenching error 
as the ratio need not converge to unity in the continuum limit.

Our final result for the decay constant is tabulated in Table~\ref{tab:result}.
To obtain the values we take the continuum extrapolation of
$f_P/\sqrt{\sigma}$ and convert it $f_P/m_\rho$ with the value of 
$\sqrt{\sigma}/m_\rho$ at $a$=0 in Fig.~\ref{fig:lattice_scale_ratio}. 
A direct continuum extrapolation of $f_P/m_\rho$ yields consistent 
results.
The quoted errors are statistical, systematic and due to scale setting
as estimated above in this order.  
We take the result for the Clover action to be our
best estimate primarily because scaling violation is smaller and also since 
the statistical ensemble is larger compared to the Wilson action.
Combining errors by quadrature we obtain the results quoted in the
abstract.

\begin{table}[tbp]
\setlength{\tabcolsep}{0.8pc}
  \begin{center}
  \caption{Results for the decay constant in MeV unit. }
    \begin{tabular}{lll}
      \hline
                        & Wilson & Clover \\
      \hline
        $f_B$           & 140(11)(15)(24) & 163(9)(8)(11)\\
        $f_{B_s}$       & 159(10)(17)(27) & 175(9)(9)(13)\\
        $f_D$           & 163(13)(18)(28) & 184(9)(9)(12)\\
        $f_{D_s}$       & 180(11)(20)(31) & 203(9)(10)(14)\\
      \hline
    \end{tabular}
  \label{tab:result}
  \end{center}
\vspace*{-9mm}
\end{table}

\vspace*{2mm}
This work is supported by the Supercomputer 
Project (No.~97-15) of High Energy Accelerator Research Organization (KEK),
and also in part by the Grants-in-Aid of 
the Ministry of Education (Nos. 08640349, 08640350, 08640404,
09246206, 09304029, 09740226).

\end{document}